\documentclass[traditabstract, twocolumns, letter, longauth]{aa}
\usepackage{txfonts}
\usepackage{graphicx}	
\usepackage{amsmath}	
\usepackage{amssymb}	
\usepackage{xcolor}     
\usepackage{soul}       
\usepackage{array}
\usepackage{mathtools}
\usepackage{layouts}
\usepackage[colorlinks=true,linkcolor=blue,citecolor=blue, urlcolor=blue]{hyperref}
\bibpunct{(}{)}{;}{a}{}{,}

\newcommand{\Jseventeen}{J1721$+$8842\xspace}

\begin{document} 

\title{
\Jseventeen: The first Einstein zig-zag lens
}
\author{
F.~Dux\thanks{Co-first authors}\inst{\ref{eso}, \ref{epfl}},  
M.~Millon$^{\star}$ \inst{\ref{stanford}, \ref{ethz}},
C.~Lemon\inst{\ref{oskarklein}},
T.~Schmidt\inst{\ref{ucla}},
F.~Courbin\inst{\ref{barca}, \ref{icrea}},
A.~J.~Shajib\inst{\ref{uchicago},\ref{kicp}}\thanks{NHFP Einstein Fellow},
T.~Treu\inst{\ref{ucla}},
S.~Birrer\inst{\ref{stonybrook}},
K.~C.~Wong\inst{\ref{utokyo}},
A.~Agnello\inst{\ref{dark}, \ref{STFC}},
A.~Andrade\inst{\ref{eso}, \ref{unab}},
A.~Galan\inst{\ref{tum},\ref{mpia}},
J.~Hjorth\inst{\ref{dark}}
E.~Paic\inst{\ref{epfl}},
S.~Schuldt\inst{\ref{milano}, \ref{inaf}},
A.~Schweinfurth\inst{\ref{mpia}, \ref{tum}},
D.~Sluse\inst{\ref{starinst}},
A.~Smette\inst{\ref{eso}},
S.~H.~Suyu\inst{\ref{tum},\ref{mpia}}
}

\institute{
    European Southern Observatory, Alonso de Córdova 3107, Vitacura, Santiago, Chile \label{eso} 
\goodbreak \and
    Institute of Physics, Laboratory of Astrophysics, Ecole Polytechnique 
    F\'ed\'erale de Lausanne (EPFL), Observatoire de Sauverny, 1290 Versoix, 
    Switzerland \label{epfl} 
\goodbreak \and 
    Kavli Institute for Particle Astrophysics and Cosmology and Department of Physics, Stanford University, Stanford, CA 94305, USA \label{stanford} 
\goodbreak \and 
    Institute for Particle Physics and Astrophysics, ETH Zurich, Wolfgang-Pauli-Strasse 27, CH-8093 Zurich, Switzerland \label{ethz}
\goodbreak \and
    Oskar Klein Centre, Department of Physics, Stockholm University, SE-106 91 Stockholm, Sweden \label{oskarklein} 
\goodbreak \and
    UCLA Physics \& Astronomy, 475 Portola Plaza, Los Angeles, CA 90095-1547, USA \label{ucla} 
\goodbreak \and
    Institut de Ciències del Cosmos, Universitat de Barcelona, Martí i Franquès, 1, E-08028 Barcelona, Spain \label{barca}  
\goodbreak \and
    ICREA, Pg. Llu\'is Companys 23, Barcelona, E-08010, Spain \label{icrea}
\goodbreak \and
    Department of Astronomy \& Astrophysics, University of Chicago, Chicago, IL 60637, USA\label{uchicago}
\goodbreak \and
    Kavli Institute for Cosmological Physics, University of Chicago, Chicago, IL 60637, USA\label{kicp}
\goodbreak \and
    Department of Physics and Astronomy, Stony Brook University, Stony Brook, NY 11794, USA\label{stonybrook}
\goodbreak \and
   Research Center for the Early Universe, Graduate School of Science, The University of Tokyo, 7-3-1 Hongo, Bunkyo-ku, Tokyo 113-0033, Japan\label{utokyo}
\goodbreak \and
 DARK, Niels Bohr Institute, University of Copenhagen, Jagtvej 155A, DK-2200 Copenhagen N, Denmark\label{dark}
 \goodbreak \and
STFC Hartree Centre, Sci-Tech Daresbury, Keckwick Lane, Daresbury, Warrington (UK) WA4 4AD \label{STFC}
\goodbreak \and
 Universidad Andres Bellos, Fernández Concha 700, Av. Las Condes alt. 13.350, Las Condes, Santiago\label{unab}
 \goodbreak \and 
   Technical University of Munich, TUM School of Natural Sciences, Physics Department, James-Franck-Straße 1, 85748 Garching, Germany\label{tum}
\goodbreak \and 
   Max-Planck-Institut für Astrophysik, Karl-Schwarzschild Straße 1, 85748 Garching, Germany \label{mpia}
\goodbreak \and
 Dipartimento di Fisica, Universit\`a  degli Studi di Milano, via Celoria 16, I-20133 Milano, Italy\label{milano}
\goodbreak \and
 INAF - IASF Milano, via A. Corti 12, I-20133 Milano, Italy\label{inaf}
\goodbreak \and
   STAR Institute, University of Li{\`e}ge, Quartier Agora, All\'ee du six Ao\^ut 19c, 4000 Li\`ege, Belgium\label{starinst}
}

\abstract{
    We report the discovery of the first example of an Einstein zig-zag lens, an extremely rare lensing configuration. In this system, \Jseventeen, six images of the same background quasar are formed by two intervening galaxies, one at redshift $z_1=0.184$ and a second one at $z_2=1.885$. Two out of the six multiple images are deflected in opposite directions as they pass the first lens galaxy on one side, and the second on the other side -- the optical paths forming zig-zags between the two deflectors. In this letter, we demonstrate that \Jseventeen, previously thought to be a lensed dual quasar, is in fact a compound lens with the more distant lens galaxy also being distorted as an arc by the foreground galaxy. 
    
    Evidence supporting this unusual lensing scenario includes: 
    1- identical light curves in all six lensed quasar images obtained from two years of monitoring at the Nordic Optical Telescope; 
    2- detection of the additional deflector at redshift $z_2=1.885$ in \textit{JWST}/NIRSpec IFU data; and 
    3- a multiple-plane lens model reproducing the observed image positions. 
    This unique configuration offers the opportunity to combine two major lensing cosmological probes: time-delay cosmography and dual source-plane lensing since \Jseventeen features multiple lensed sources forming two distinct Einstein radii of different sizes, one of which being a variable quasar. We expect tight constraints on $H_0$ and $w$ by combining these two probes on the same system. The $z_2=1.885$ deflector, a quiescent galaxy, is also the highest-redshift strong galaxy-scale lens with a spectroscopic redshift measurement.
}

\titlerunning{\Jseventeen : A first Einstein zig-zag lens}
\authorrunning{Dux F., Millon M., Lemon C., et al.}
\maketitle

\section{Introduction}
\label{sec:introduction}

Strong gravitational lensing is a powerful tool to measure distances and ratios of distances in the Universe~\citep{Refsdal1964}. 
To date, two main methods have provided the most stringent constraints on cosmological parameters. The first, time-delay cosmography, uses the time delay between the arrival of different images of a variable lensed source (typically quasars or supernovae) to infer the \emph{time-delay distance}, i.e. a ratio of distances, to the lens, to the source, and between the lens and the source \citep[see e.g.][for a review]{Birrer2022}. 
From these time-delay distance measurements, the COSMOGRAIL, H0LiCOW, SHARP, STRIDES, and TDCOSMO collaborations have achieved a 2 to 8\% measurement of the Hubble Constant ($H_0$), depending on the assumptions made to describe the mass profile of the deflector galaxy \citep{Chen2019, Wong2020, Birrer2020, Millon2020a, Shajib2020}. 
The first discovered cases of supernovae lensed by galaxy clusters also allowed a 
6\%~\citep{Kelly2023a, Grillo2024},
and an 11\%~\citep{Pascal2024}
determination of $H_0$, 
while the search for a galaxy-scale lens of a supernova suitable for time-delay cosmography is still ongoing.

The second method uses double source-plane lenses (DSPL). 
This method leverages rare lensing configurations where two background objects are lensed by the same foreground galaxy, forming two Einstein rings of different radii \citep{Gavazzi2008}. 
The ratio of the two radii is proportional to ratios involving distances to the two sources and between the lens and the sources \citep[see e.g.][for details]{Collett2012}. 
These ratios of distances are not sensitive to $H_0$ but constrain the \emph{expansion history of the Universe} and, thus, can provide competitive measurements of the matter density parameter, $\Omega_m$, and of the dark energy equation of state parameter, $w$ \citep{Collett2014, Linder2016}. Although the presence of multiple background sources is common for galaxy clusters, this information is needed to infer their complex mass distribution, which reduces the precision on the cosmological parameters \citep[e.g.][]{Jullo2010, Caminha2022}. In contrast, galaxy-scale lenses tend to be very well fitted with simple mass distributions \citep[e.g.][]{Shajib2021,Sheu2024} but galaxy-scale DSPL are unfortunately extremely rare. Only six systems have been spectroscopically confirmed to date  \citep{Gavazzi2008, Tu2009, Tanaka2016, Schuldt2019, Lemon2023, Bolamperti2023}, although several dozen should be found in wide-field surveys such as LSST or Euclid \citep{Sharma2023}. Until more galaxy-scale DSPL are found, this probe will remain limited by statistical errors due to the small sample size. 

\begin{figure}[t!]
    \centering
    \includegraphics[width=0.95\columnwidth]{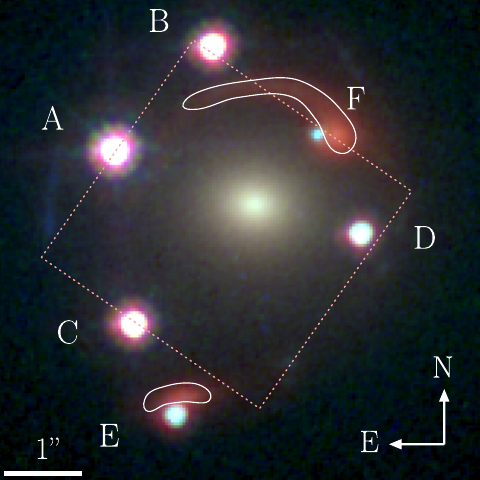}
    \caption[Caption with footnote]{
    \textit{HST}/WFC3%
    \footnotemark 
    composite image of \Jseventeen. The six lensed point images of the background quasar are labeled with letters. The red lensed arc, previously believed to be the host galaxy of a second source quasar, is highlighted in white. This arc is actually an additional deflector at an intermediate redshift, which is itself lensed by the central foreground galaxy in the image. The dotted red square indicates the footprint of the \textit{JWST}/NIRSpec observation which enabled the measurement of the arc's redshift.
    }\label{fig:hst_composite_annotated}
\end{figure}

Here, we present a unique case of galaxy-scale strong lensing, \Jseventeen, originally discovered as a quadruply-imaged source at redshift $z_s = 2.38$ by \citet{Lemon2018} using \textit{Gaia} and Pan-STARRS, which was subsequently interpreted as a lensed dual active galactic nucleus (dual-AGN) in \cite{Lemon2022} and \cite{Mangat2021} due to the presence of two additional images in follow-up data (labeled E and F in Fig.~\ref{fig:hst_composite_annotated}). 
We demonstrate in this letter that these six images are all from the the same background quasar, 
as made possible by another deflector at redshift $z_2 = 1.885$ in addition to the known lens at redshift $z_1 = 0.184$. 
This second deflector, contoured in white in Fig.~\ref{fig:hst_composite_annotated}, was previously mistaken for a source galaxy at the same redshift as the quasar, from which the black holes may have been ejected in a post-merger scenario \citep{Mangat2021}.
Since the more distant deflector is also lensed into a partial Einstein ring, \Jseventeen qualifies as both a DSPL and a time-delay lens. 
This exceptional system provides a unique opportunity to combine the two major strong lensing probes mentioned above. 

In this letter, we present evidence from the light curves obtained at the Nordic Optical Telescope (NOT), new redshift measurements from the James Web Space Telescope (\textit{JWST}) Near InfraRed Spectrograph (NIRSpec), and updated lens models, which unambiguously confirm the scenario where one single source is lensed in \Jseventeen. 
Full lens models, time-delay measurements and cosmology constraints derived from this system will be published in follow-up papers as part of the TDCOSMO collaboration (Schmidt et al. in prep., Dux et al. in prep., Millon et al. in prep.). 

Throughout this work, we adopt a flat $\Lambda$CDM cosmology with $H_0=70{\rm \,km \,s^{-1}\, Mpc^{-1}}$ and $\Omega_m=0.3$.

\footnotetext{PI Tommaso Treu, \textit{HST} proposal ID 15652} 

\section{Evidence for single source lensing}

\subsection{Identical light curves}
\begin{figure*}
    \centering
    \includegraphics[width=\textwidth]{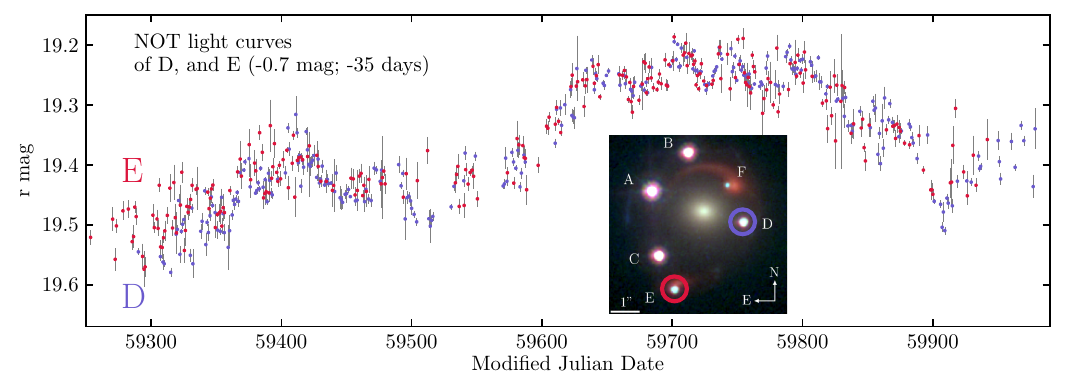}\vspace{-0.3cm}
    \includegraphics[width=0.975\textwidth]{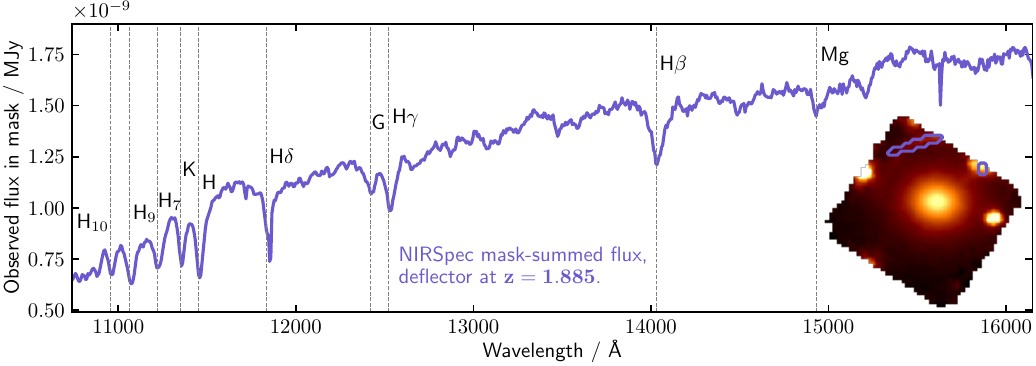}
    \vspace{-0.2cm}
    \caption{Summary of evidence showing the unique source and double lens nature of \Jseventeen. \textit{Top:} NOT light curves of the D and E lensed images of \Jseventeen. The inset is a color-composite \textit{HST}/WFC3 image of the system
    with the labeling of the lensed components.
    The two light curves are identical within their noise envelope, with D preceding E by about 35 days. 
    \textit{Bottom:} \textit{JWST}/NIRSpec spectrum of the far deflector arced by the foreground lens.
    The spectrum was extracted from the mask drawn on the inset, which itself is a single slice of the NIRSpec cube. The redshift was determined from the overlaid absorption lines. 
    Note that the total flux presented here is not corrected for the magnification due to the lensing.}
    \label{fig:lc_spec}
\end{figure*}

\Jseventeen was imaged daily in the $r$ band for two full years at the NOT, from February 2021 to August 2023. 
It is circumpolar, allowing for continuous monitoring at the latitude of NOT.
The general observation strategy and data reduction will be described in an upcoming paper, together with more targets monitored at the same facility.
After the completion of the observations, the suspicion that \Jseventeen is in fact a single quasar, arose from the light curves of the  E and F lensed images, which match those of A, B, C and D (see Fig.~\ref{fig:hst_composite_annotated} for the labeling of the lensed images).
For illustration, the D and E curves, shown in Fig.~\ref{fig:lc_spec}, display the same variations with a $\sim$35-day offset. 
This non-zero delay rules out extraction systematics, which would cause simultaneous variations. 
This similarity contradicts the dual quasar hypothesis, in which the curves would be uncorrelated:
if two quasars were present in the source plane, they would be separated by at least 6\,kpc \citep{Lemon2022}, and thus it is impossible that one AGN would trigger a correlated emission in the second one.

However, there might remain a possibility that two independently varying quasars could show variations matching within their noise envelopes, even though the odds seem slim over a long baseline.
To assess these odds, we generated a million mocks for which we calculated the best correlation with the D curve with time shifts between $-$100 and +100 days. 
In doing so, we assumed that the variability of both quasars could be modeled with Ornstein-Uhlenbeck (OU) processes, the parameters of which (correlation time $\tau$ and volatility $\sigma$) were fitted on the D curve.
The best attained correlation was 0.91, translating to a resemblance of low-frequency features, whereas D and E have a correlation of 0.95, with even high-frequency features matching.
Thus, we failed to find a properly matching curve in one million draws. 
To assess how many mocks it would take to match even the higher-frequency features, 
we considered two identical OU processes initially differing by a value within the noise envelope $\sigma_{m}$. 
By analyzing the difference between their states after a time step $\Delta t$, we determined that this difference follows a Gaussian distribution with zero mean and variance $2\sigma^{2}\Delta t + (\Delta t/\tau)^{2}\sigma_{m}^{2}$. 
Using the cumulative distribution function of the normal distribution, $\Phi$, we calculated the probability that the two processes remain within the noise envelope after $\Delta t$ days, given in Eq.~(\ref{eq:uoodds}) below.
\begin{equation}
\mathcal{P}=\frac{1}{2\Phi(1)-1} 
\left[
2\Phi\left(\frac{\sigma_{m}}{\sqrt{2(\Delta t\sigma^{2}+\left(\frac{\Delta t}{\tau}\right)^2\sigma_{m}^{2})}}\right)-1
\right]
\label{eq:uoodds}
\end{equation}
We multiply this probability for a succession of time lags  $\Delta t$, until we reach the length of the light curves. 
We take $\Delta t \sim$~40~days as the size of the smallest clearly captured features, and we obtain the probability of observing two identical light curves from two independent quasars over this baseline to be about $10^{-12}$.
This is assuming the $\Delta t \sim -35$ days delay required for matching the two curves, but the existence of a delay is expected regardless of the scenario, with moreover all six curves matching in a similar way with time shifts only.
Furthermore, this all neglects the odds of two independent quasars have the same variability parameters. 
Overall, both numerical and analytical approaches indicate vanishing odds of two quasars producing light curves this similar over the 700 days monitoring baseline.

\subsection{Redshift of the red lensed arc}

\citet{Lemon2018} first confirmed the four bright images (A, B, C, and D) to be at a redshift of 2.38, and subsequent long-slit spectroscopy confirmed image E to be at the same redshift, though the signal-to-noise ratio did not allow for any clear comparison of the spectral features \citep{Lemon2022}. 
Image F was not directly measured, but was assumed to be the counter-image of E. 
We note that these analyses were done solely on ground-based imaging, and thus the arc seen in subsequent Hubble Space Telescope (\textit{HST}) imaging (contoured in white in Fig.~\ref{fig:hst_composite_annotated}) was not used to conclude the dual-source nature of the system. 
\citet{Mangat2021} did use the \textit{HST} imaging and suggested that the arc was another source galaxy, located at the same redshift as the quasar. 
These authors noticed that the center position of this galaxy did not match with the position of any of the quasars and proposed the hypothesis that the quasars were recoiling black holes, ejected from their host galaxy after a merger.   
No redshift for the arc had been reported, and all previous lens modeling assumed a single lens plane.

Thanks to a recent, deep \textit{JWST}/NIRSpec \citep{Jakobsen2022} observation\footnote{
PI Tommaso Treu and Anowar J. Shajib, \textit{JWST} proposal ID 2974
},
this hypothesis could finally be tested. 
The footprint of the \textit{JWST}/NIRSpec observation is shown in Fig.~\ref{fig:hst_composite_annotated}, with a part of the red arc falling into the field of view.
The extracted spectrum of the arc, and its extraction aperture on a slice of the NIRSpec cube, are displayed in the bottom part of Fig.~\ref{fig:lc_spec}. 
The spectrum features Balmer absorption lines that make a redshift determination straightforward: $z_{2}=1.885\pm0.001$\footnote{
Barycentric frame, determined from the Balmer and calcium lines.
}. 
This is incompatible with the redshift of the background quasar, which can also be determined from the NIRSpec data: $z_{\rm F}\sim 2.382\pm0.002$, in agreement with the results of~\cite{Lemon2022}.
Therefore, rather than being the host galaxy of a second quasar, 
the arc is due to a separate galaxy, lensed by the foreground deflector at $z_1=0.184$, and itself acting as a lens for the quasar light.
At redshift $z_2=1.885$, this is the highest redshift spectroscopically-confirmed galaxy-scale lens known, breaking the previous record of $z=1.62$, for a galaxy embedded in a cluster \citep{Wong2014}. Its redshift is also comparable to that of the lens found by \cite{vanDokkum2023}, although only a photometric redshift \citep[$z=2.02 \pm 0.02$, ][]{Mercier2024} has been obtained for this system.

\begin{figure*}
    \centering
    \includegraphics[width=0.50\textwidth]{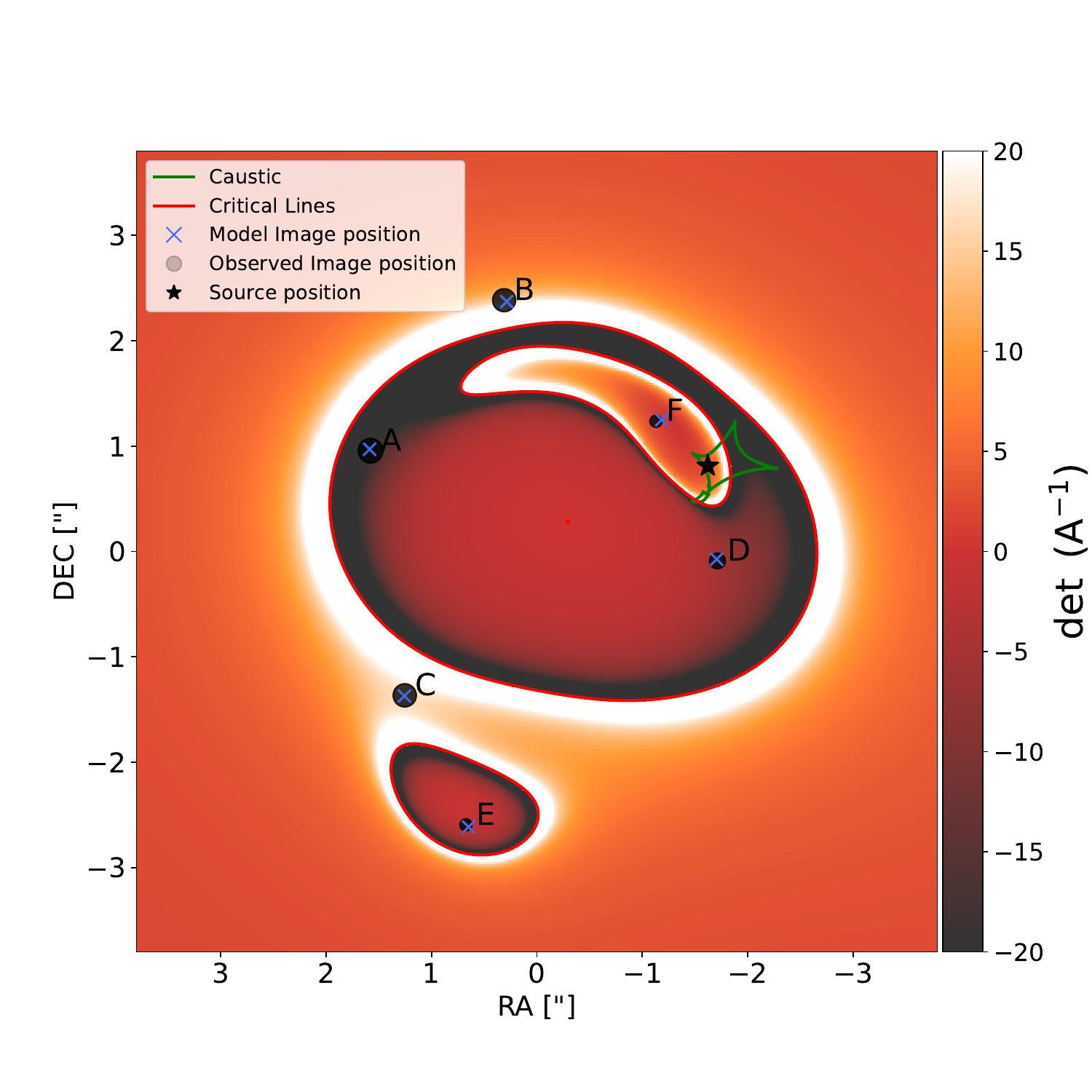}
    \hspace{1cm}
    \includegraphics[width=0.43\linewidth]{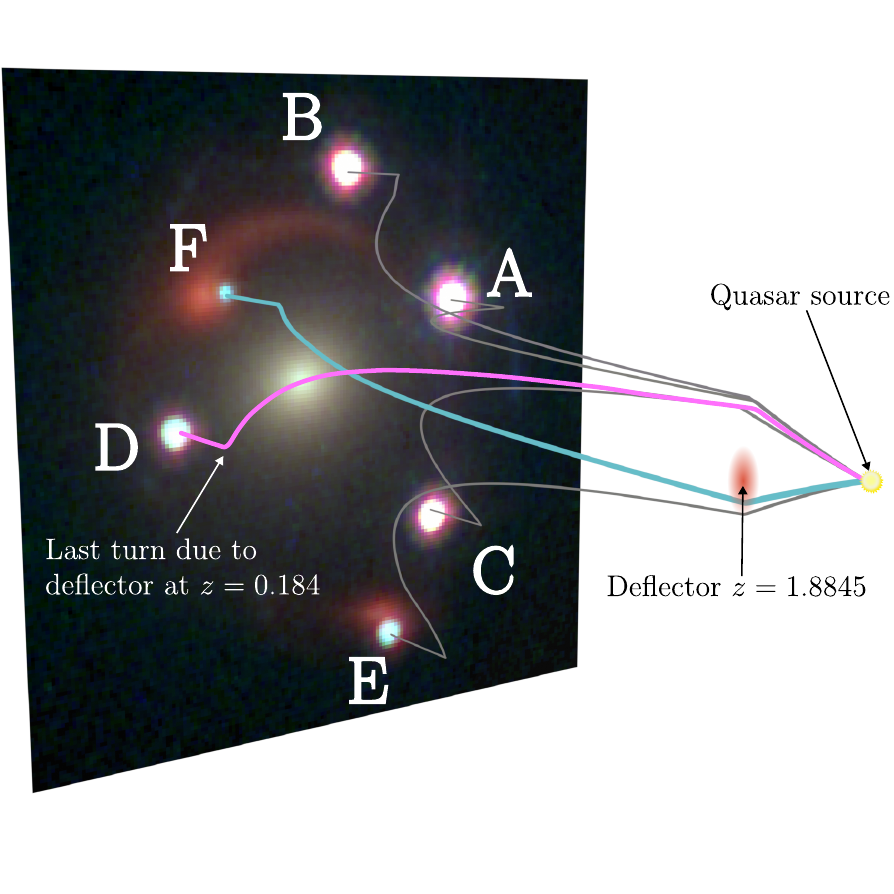}
    \vspace{-0.5cm}
    \caption{
    \textit{Left:} Magnification map. Critical lines in the lens plane are shown in red, while the corresponding inner caustic in the source plane are represented in green. The predicted positions of the images from our mode (black circles with size proportional to the image's magnification) reproduce the observed image positions (blue crosses). The quasar's position in the source plane is indicated with a star. 
    \textit{Right:}
    Visualization of the optical paths of the lensed images.
    The two zig-zag paths, D and F, are marked in pink and blue respectively.
    The deflectors cause two sets of sharp turns, while the smooth curvature seen in all paths is due to the expansion of space.
    }\label{fig:lens_model}
\end{figure*}

\subsection{Lens model} \label{lensmodels}

Original models published by \cite{Schmidt2023} and \cite{Ertl2023} assumed two quasars in the source plane, and modeled the lensed arcs as one of the quasar host galaxies at redshift $z_s = 2.38$. However, \cite{Schmidt2023} observed some unexpectedly large astrometric errors for this system whereas \cite{Ertl2023} noted an offset between the quasar position and its host galaxy. This offset was also noticed in the models by \cite{Mangat2021}. In light of the new redshift measurement from the \textit{JWST}/NIRSpec observations, we updated the lenstronomy~\citep{Birrer2018, Birrer2021} model of \cite{Schmidt2023}, adding this additional perturber at redshift $z_2=1.885$. 
The updated multi-plane lens model reproduces the positions of the six lensed images, matching the astrometry measured from \textit{HST} imaging with a root-mean-squared error of $0.03''$. 
The model includes a power-Law elliptical mass distribution (PEMD) at $z_1=0.184$ to describe the foreground lensing galaxy, a singular isothermal sphere (SIS) at $z_2=1.885$, and external shear. Additionally, we include four galaxies located between 10\arcsec and 33\arcsec\ from the optical axis that are visible in the \textit{HST} images. These perturbers do not have spectroscopic redshift measurements but, given their color in the \textit{HST} images, we assumed that they are located in the  $z_1=0.184$ lens plane. 
They are modeled as SIS profiles and have little impact on the total convergence.
In Fig.~\ref{fig:lens_model}, we show the magnification map of our lens model, alongside the predicted and observed image positions and the optical paths taken by each lensed image. Both the ABCD quartet and the EF doublet are initially deflected by the $z_2=1.885$ deflector, yet they still form two separate and compact groups of light rays at this redshift.
Moreover, the two groups are separated by $0.7''$ , which translates to a separation of $6$\,kpc, matching the reported distance between the two initially postulated  quasar sources. 
Thus, not only does the model reproduce the observational data, but it also qualitatively explains how the former hypothesis of two sources could be fitted so well with a single deflector.

\section{Discussion}
To our knowledge, the formation of six multiple images of the same background source has never been observed in galaxy-scale strong lenses. 
However, this lensing configuration was originally studied by \cite{Kochanek1988} and \cite{Erdl1993} \citep[see also][]{Moller2001, Werner2008}. More recent studies by \cite{Collett2016} predicted that such system could form \emph{``only if the second lens is multiply imaged by the first and the Einstein radius of the second lens is comparable to, but does not exceed that of the first''}. \Jseventeen fulfills these two conditions, as it is clear from the \textit{HST} images that the most distant deflector is doubly imaged, with the main arc visible in the vicinity of image F and a counter-image close to image E. In addition, the Einstein radius of this deflector, inferred from our lens model, is $\theta_{E,2} = 0.359 \pm 0.014"$ translating to a total enclosed mass of $M_2 = (2.31 \pm 0.19) \cdot 10^{11}M_{\odot}$. 
This makes the more distant deflector slightly less massive than the foreground one, whose Einstein radius is $\theta_{E,1} = 1.744 \pm 0.013"$, corresponding to a total enclosed mass of $M_1 = (2.74 \pm 0.04) \cdot 10^{11}M_{\odot}$.

The ray tracing of the multiple images, illustrated in the right panel of Figure \ref{fig:lens_model}, reveals the zig-zag formed by the two deflectors: image F passes by the far deflector on one side, then is refocused by the foreground galaxy as it passes it on its other side. 
Image D also experiences two deflections in opposite directions. 

Additionally, the DPSL nature of this system, with two source planes and six measured time delays, might also partially lift the Mass-Sheet Degeneracy~\citep[MSD, ][]{Falco1985, Schneider2013}, which currently is the main source of uncertainties in time-delay cosmography analyses. 
In recent works \citep[e.g.][]{Birrer2020, Shajib2023}, a distinction has been made between the \emph{external} MSD, attributed to the net lensing effect of line-of-sight structures around the main lens -- constrained by galaxy number counts \citep[e.g.][]{Rusu2017, Wells2023} or weak lensing analysis \citep{Tihhonova2018,Tihhonova2020} -- and the \emph{internal} MSD, interpreted as a subtle change of slope in the mass profile of the lensing galaxy, possibly due to the presence of a cored dark matter component \citep[see][for details]{Blum2020, Blum2024}. 
The \emph{internal} MSD parameter $\lambda_{int}$, can be constrained with a measurement of the stellar kinematics of the lensing galaxy, ideally spatially resolved, which provides a second estimate of its mass profile. 
Although the MSD can be generalised for DSPLs \citep[][]{Schneider2014a,Schneider2014b}, it requires a very specific transformation in both lens planes to maintain an exact degeneracy leaving the lensing observables unchanged. 
According to \cite{Schneider2014a}, the effect of a cored profile for the first deflector ($\lambda_{int,1}$ < 1) $-$ equivalent to a positive convergence mass sheet $-$ requires a mass sheet in the second lens plane with negative convergence ($\lambda_{int,2}$ > 1) to rescale only the time delays, and thus $H_0$, while leaving other lensing observables unaffected by this transformation. 
Although this is mathematically possible, such a combination of positive and negative mass sheets lacks a clear physical interpretation, as the model proposed by \cite{Blum2024} would require some very specific fine-tuning of its redshift dependence to produce alternating cored and cusp profiles.
The hypothesis made by recent hierarchical Bayesian analysis \citep{Birrer2020} that a population of lenses should share common properties for their mass profile \citep[i.e., $\lambda_{int}$ is drawn from a parent distribution common to all lensing galaxies,][]{Birrer2020} might also break the double-source plane equivalent of the MSD. 
Two mass sheets with opposite sign would be disfavored under the aforementioned hypothesis. 
Whether all the available data can constrain both $H_0$ and $w$ simultaneously, even when considering maximally degenerate lensed models is left for the future modeling and cosmology papers on this lens.

\citet{Collett2016} calculated the optical depth for such compound lensing events for varying source redshifts and for each number of images, finding the probability of six-image systems at $ z > 2$ to be $\sim10^{-8}$. This means that such Einstein zig-zag lens occurs once in 100 million line-of-sights. Comparing to the approximate optical depth of multiple lensing of quasars -- $\sim 10^{-3.3}$ found by systematic lensed quasar searches \citep{oguri2012} -- suggests that 1 in 50$\,$000 lensed quasars should be a zig-zag lens, however only $\sim300$ such systems are known \citep{Lemon2023}. This naive approximation does not account for magnification bias, and further if we consider the optical depth to a compound lens with any number of images, the probability increases by a factor of 100. Given that recent large sky-surveys has now imaged several billions of galaxies, it is not surprising that such a system has been found, especially when considering that the high magnification of the some of the multiple images (ranging from 14 to 19 for image A, B and C) have made its detection easier.

\section{Conclusion}
We have shown that \Jseventeen is not a lensed dual quasar as previously suggested, 
but rather a single quasar imaged six times by the combined gravitational deflections of two galaxies: one at $z_1=0.184$, and another at $z_2=1.885$. 
This scenario is consistent with all available data: identical light curves in all lensed images, a lens model that reproduces the \textit{HST} images and the detection of an additional deflector, itself lensed, from  \textit{JWST}/NIRSpec spectroscopy. 
This image configuration is extremely rare, with the optical path of two images, namely D and F, forming a zig-zag between the two deflectors. This is the first known example of an Einstein zig-zag lens.

The source quasar of \Jseventeen has a Proximate Damped Lyman-$\alpha$ (PDLA) system, which only about 1 in 3000 quasars have~\citep{Finley2013, Lemon2022}. This characteristic, combined with its exceptional lensing configuration, will allow us to probe the PDLA along six different line-of-sight, offering the possibility to study these rare systems on sub-parsec scales. 

In addition, we report that the lens at $z_2=1.885$ is quiescent: its spectrum shows the presence of neutral hydrogen gas, but no emission lines typical of star formation. 
This deflector is similar in nature and redshift to that found by \citet{vanDokkum2023} and \cite{Mercier2024}. To date, this is also the highest redshift strong galaxy lens confirmed with spectroscopy. A precise mass model of \Jseventeen can answer questions about the compactness of this $z_2=1.885$ lens, its dark matter fraction, and other questions studied by \citet{vanDokkum2023}.

Finally, the source galaxy of \Jseventeen hosts a variable AGN and it is the most Northern gravitational lens known, which makes continuous monitoring from the ground possible. 
This has made the time-delay measurements between all six images possible. 
This system is therefore valuable for future $H_0$ measurement through time-delay cosmography, where high-redshift lenses are known to provide tighter constraints on the cosmological parameters \citep{Paraficz2009}. 
Its DSPL nature is also advantageous: its lens model can also constrain ratio of distances between the observer, the lens and the two sources, allowing precise measure of the expansion history of the Universe, with particular sensitivity to the dark energy equation of state. 
By combining these two strong lensing probes on the same system, it will be possible to probe $H_0$, $w$, and the mass profile of the higher redshift deflector, which is of interest for the population of massive galaxies at $z\sim2$.

\begin{acknowledgements}
The first two authors should be regarded as joint first authors.
MM acknowledges support by the SNSF (Swiss National Science Foundation) through mobility grant P500PT\_203114 and return CH grant P5R5PT\_225598.
FC acknowledges support from the SNSF and by the European Research Council (ERC) under the European Unions Horizon 2020 research and innovation programme 
(COSMICLENS: grant agreement No 787886). This project has received funding from the European Union’s Horizon Europe research and innovation programme under the Marie Sklodovska-Curie grant agreement No 101105725.
SS has received funding from the European Union’s Horizon 2022 research and innovation programme under the Marie Skłodowska-Curie grant agreement No 101105167 — FASTIDIoUS.
SHS thanks the Max Planck Society for support through the Max Planck Fellowship.  This research is supported in part by the Excellence Cluster ORIGINS which is funded by the Deutsche Forschungsgemeinschaft (DFG, German Research Foundation) under Germany's Excellence Strategy -- EXC-2094 -- 390783311.
This work is supported by JSPS KAKENHI Grant Numbers JP24K07089, JP24H00221.AA has been partially supported by the Villum Experiment grant \textit{Cosmic Beacons} (project number 36225; PIs Agnello, Izzo).
This work is based in part on observations made with the NASA/ESA/CSA James Webb Space Telescope. The data were obtained from the Mikulski Archive for Space Telescopes at the Space Telescope Science Institute, which is operated by the Association of Universities for Research in Astronomy, Inc., under NASA contract NAS 5-03127 for JWST. These observations are associated with program \#2974. Support for program \#2974 was provided by NASA through a grant from the Space Telescope Science Institute, which is operated by the Association of Universities for Research in Astronomy, Inc., under NASA contract NAS 5-03127. This work was supported by research grants (VIL16599, VIL54489) from VILLUM FONDEN.
The NOT light curves were extracted with \texttt{lightcurver}~\citep{lightcurver}.
\end{acknowledgements}
\newpage
\bibliographystyle{aa}
\bibliography{biblio}

\begin{thebibliography}{55}
\expandafter\ifx\csname natexlab\endcsname\relax\def\natexlab#1{#1}\fi

\bibitem[{Birrer \& Amara(2018)}]{Birrer2018}
Birrer, S. \& Amara, A. 2018, Physics of the Dark Universe, 22, 189–201

\bibitem[{{Birrer} {et~al.}(2024){Birrer}, {Millon}, {Sluse}, {Shajib},
  {Courbin}, {Erickson}, {Koopmans}, {Suyu}, \& {Treu}}]{Birrer2022}
{Birrer}, S., {Millon}, M., {Sluse}, D., {et~al.} 2024, \ssr, 220, 48

\bibitem[{{Birrer} {et~al.}(2021){Birrer}, {Shajib}, {Gilman}, {Galan},
  {Aalbers}, {Millon}, {Morgan}, {Pagano}, {Park}, {Teodori}, {Tessore},
  {Ueland}, {Van de Vyvere}, {Wagner-Carena}, {Wempe}, {Yang}, {Ding},
  {Schmidt}, {Sluse}, {Zhang}, \& {Amara}}]{Birrer2021}
{Birrer}, S., {Shajib}, A., {Gilman}, D., {et~al.} 2021, The Journal of Open
  Source Software, 6, 3283

\bibitem[{{Birrer} {et~al.}(2020){Birrer}, {Shajib}, {Galan}, {Millon}, {Treu},
  {Agnello}, {Auger}, {Chen}, {Christensen}, {Collett}, {Courbin}, {Fassnacht},
  {Koopmans}, {Marshall}, {Park}, {Rusu}, {Sluse}, {Spiniello}, {Suyu},
  {Wagner-Carena}, {Wong}, {Barnab{\`e}}, {Bolton}, {Czoske}, {Ding},
  {Frieman}, \& {Van de Vyvere}}]{Birrer2020}
{Birrer}, S., {Shajib}, A.~J., {Galan}, A., {et~al.} 2020, \aap, 643, A165

\bibitem[{{Blum} {et~al.}(2020){Blum}, {Castorina}, \&
  {Simonovi{\'c}}}]{Blum2020}
{Blum}, K., {Castorina}, E., \& {Simonovi{\'c}}, M. 2020, \apjl, 892, L27

\bibitem[{{Blum} \& {Teodori}(2024)}]{Blum2024}
{Blum}, K. \& {Teodori}, L. 2024, arXiv e-prints, arXiv:2409.04134

\bibitem[{{Bolamperti} {et~al.}(2023){Bolamperti}, {Grillo}, {Ca{\~n}ameras},
  {Suyu}, \& {Christensen}}]{Bolamperti2023}
{Bolamperti}, A., {Grillo}, C., {Ca{\~n}ameras}, R., {Suyu}, S.~H., \&
  {Christensen}, L. 2023, \aap, 671, A60

\bibitem[{{Caminha} {et~al.}(2022){Caminha}, {Suyu}, {Grillo}, \&
  {Rosati}}]{Caminha2022}
{Caminha}, G.~B., {Suyu}, S.~H., {Grillo}, C., \& {Rosati}, P. 2022, \aap, 657,
  A83

\bibitem[{{Chen} {et~al.}(2019){Chen}, {Fassnacht}, {Suyu}, {Rusu}, {Chan},
  {Wong}, {Auger}, {Hilbert}, {Bonvin}, {Birrer}, {Millon}, {Koopmans},
  {Lagattuta}, {McKean}, {Vegetti}, {Courbin}, {Ding}, {Halkola}, {Jee},
  {Shajib}, {Sluse}, {Sonnenfeld}, \& {Treu}}]{Chen2019}
{Chen}, G. C.~F., {Fassnacht}, C.~D., {Suyu}, S.~H., {et~al.} 2019, \mnras,
  490, 1743

\bibitem[{{Collett} \& {Auger}(2014)}]{Collett2014}
{Collett}, T.~E. \& {Auger}, M.~W. 2014, \mnras, 443, 969

\bibitem[{{Collett} {et~al.}(2012){Collett}, {Auger}, {Belokurov}, {Marshall},
  \& {Hall}}]{Collett2012}
{Collett}, T.~E., {Auger}, M.~W., {Belokurov}, V., {Marshall}, P.~J., \&
  {Hall}, A.~C. 2012, \mnras, 424, 2864

\bibitem[{{Collett} \& {Bacon}(2016)}]{Collett2016}
{Collett}, T.~E. \& {Bacon}, D.~J. 2016, \mnras, 456, 2210

\bibitem[{Dux(2024)}]{lightcurver}
Dux, F. 2024, Journal of Open Source Software, 9, 6775

\bibitem[{{Erdl} \& {Schneider}(1993)}]{Erdl1993}
{Erdl}, H. \& {Schneider}, P. 1993, \aap, 268, 453

\bibitem[{{Ertl} {et~al.}(2023){Ertl}, {Schuldt, S.}, {Suyu, S. H.}, {Schmidt,
  T.}, {Treu, T.}, {Birrer, S.}, {Shajib, A. J.}, \& {Sluse, D.}}]{Ertl2023}
{Ertl}, S., {Schuldt, S.}, {Suyu, S. H.}, {et~al.} 2023, \aap, 672, A2

\bibitem[{{Falco} {et~al.}(1985){Falco}, {Gorenstein}, \&
  {Shapiro}}]{Falco1985}
{Falco}, E.~E., {Gorenstein}, M.~V., \& {Shapiro}, I.~I. 1985, \apjl, 289, L1

\bibitem[{{Finley} {et~al.}(2013){Finley}, {Petitjean}, {P{\^a}ris},
  {Noterdaeme}, {Brinkmann}, {Myers}, {Ross}, {Schneider}, {Bizyaev},
  {Brewington}, {Ebelke}, {Malanushenko}, {Malanushenko}, {Oravetz}, {Pan},
  {Simmons}, \& {Snedden}}]{Finley2013}
{Finley}, H., {Petitjean}, P., {P{\^a}ris}, I., {et~al.} 2013, \aap, 558, A111

\bibitem[{{Gavazzi} {et~al.}(2008){Gavazzi}, {Treu}, {Koopmans}, {Bolton},
  {Moustakas}, {Burles}, \& {Marshall}}]{Gavazzi2008}
{Gavazzi}, R., {Treu}, T., {Koopmans}, L. V.~E., {et~al.} 2008, \apj, 677, 1046

\bibitem[{{Grillo} {et~al.}(2024){Grillo}, {Pagano}, {Rosati}, \&
  {Suyu}}]{Grillo2024}
{Grillo}, C., {Pagano}, L., {Rosati}, P., \& {Suyu}, S.~H. 2024, \aap, 684, L23

\bibitem[{{Jakobsen} {et~al.}(2022){Jakobsen}, {Ferruit}, {Alves de Oliveira},
  {Arribas}, {Bagnasco}, {Barho}, {Beck}, {Birkmann}, {B{\"o}ker}, {Bunker},
  {Charlot}, {de Jong}, {de Marchi}, {Ehrenwinkler}, {Falcolini}, {Fels},
  {Franx}, {Franz}, {Funke}, {Giardino}, {Gnata}, {Holota}, {Honnen}, {Jensen},
  {Jentsch}, {Johnson}, {Jollet}, {Karl}, {Kling}, {K{\"o}hler}, {Kolm},
  {Kumari}, {Lander}, {Lemke}, {L{\'o}pez-Caniego}, {L{\"u}tzgendorf},
  {Maiolino}, {Manjavacas}, {Marston}, {Maschmann}, {Maurer}, {Messerschmidt},
  {Moseley}, {Mosner}, {Mott}, {Muzerolle}, {Pirzkal}, {Pittet}, {Plitzke},
  {Posselt}, {Rapp}, {Rauscher}, {Rawle}, {Rix}, {R{\"o}del}, {Rumler},
  {Sabbi}, {Salvignol}, {Schmid}, {Sirianni}, {Smith}, {Strada}, {te Plate},
  {Valenti}, {Wettemann}, {Wiehe}, {Wiesmayer}, {Willott}, {Wright}, {Zeidler},
  \& {Zincke}}]{Jakobsen2022}
{Jakobsen}, P., {Ferruit}, P., {Alves de Oliveira}, C., {et~al.} 2022, \aap,
  661, A80

\bibitem[{{Jullo} {et~al.}(2010){Jullo}, {Natarajan}, {Kneib}, {D'Aloisio},
  {Limousin}, {Richard}, \& {Schimd}}]{Jullo2010}
{Jullo}, E., {Natarajan}, P., {Kneib}, J.~P., {et~al.} 2010, Science, 329, 924

\bibitem[{{Kelly} {et~al.}(2023){Kelly}, {Rodney}, {Treu}, {Oguri}, {Chen},
  {Zitrin}, {Birrer}, {Bonvin}, {Dessart}, {Diego}, {Filippenko}, {Foley},
  {Gilman}, {Hjorth}, {Jauzac}, {Mandel}, {Millon}, {Pierel}, {Sharon},
  {Thorp}, {Williams}, {Broadhurst}, {Dressler}, {Graur}, {Jha}, {McCully},
  {Postman}, {Schmidt}, {Tucker}, \& {von der Linden}}]{Kelly2023a}
{Kelly}, P.~L., {Rodney}, S., {Treu}, T., {et~al.} 2023, Science, 380, abh1322

\bibitem[{{Kochanek} \& {Apostolakis}(1988)}]{Kochanek1988}
{Kochanek}, C.~S. \& {Apostolakis}, J. 1988, \mnras, 235, 1073

\bibitem[{{Lemon} {et~al.}(2023){Lemon}, {Anguita}, {Auger-Williams},
  {Courbin}, {Galan}, {McMahon}, {Neira}, {Oguri}, {Schechter}, {Shajib},
  {Treu}, {Agnello}, \& {Spiniello}}]{Lemon2023}
{Lemon}, C., {Anguita}, T., {Auger-Williams}, M.~W., {et~al.} 2023, \mnras,
  520, 3305

\bibitem[{{Lemon} {et~al.}(2022){Lemon}, {Millon}, {Sluse}, {Courbin}, {Auger},
  {Chan}, {Paic}, \& {Agnello}}]{Lemon2022}
{Lemon}, C., {Millon}, M., {Sluse}, D., {et~al.} 2022, \aap, 657, A113

\bibitem[{{Lemon} {et~al.}(2018){Lemon}, {Auger}, {McMahon}, \&
  {Ostrovski}}]{Lemon2018}
{Lemon}, C.~A., {Auger}, M.~W., {McMahon}, R.~G., \& {Ostrovski}, F. 2018,
  \mnras, 479, 5060

\bibitem[{{Linder}(2016)}]{Linder2016}
{Linder}, E.~V. 2016, \prd, 94, 083510

\bibitem[{{Mangat} {et~al.}(2021){Mangat}, {McKean}, {Brilenkov}, {Hartley},
  {Stacey}, {Vegetti}, \& {Wen}}]{Mangat2021}
{Mangat}, C.~S., {McKean}, J.~P., {Brilenkov}, R., {et~al.} 2021, \mnras, 508,
  L64

\bibitem[{{Mercier} {et~al.}(2024){Mercier}, {Shuntov}, {Gavazzi},
  {Nightingale}, {Arango}, {Ilbert}, {Amvrosiadis}, {Ciesla}, {Casey}, {Jin},
  {Faisst}, {Andika}, {Drakos}, {Enia}, {Franco}, {Gillman}, {Gozaliasl},
  {Hayward}, {Huertas-Company}, {Kartaltepe}, {Koekemoer}, {Laigle}, {Le
  Borgne}, {Magdis}, {Mahler}, {Maraston}, {Martin}, {Massey}, {McCracken},
  {Moutard}, {Paquereau}, {Rhodes}, {Robertson}, {Sanders}, {Toft},
  {Trebitsch}, {Tresse}, \& {Vijayan}}]{Mercier2024}
{Mercier}, W., {Shuntov}, M., {Gavazzi}, R., {et~al.} 2024, \aap, 687, A61

\bibitem[{{Millon} {et~al.}(2020){Millon}, {Galan}, {Courbin}, {Treu}, {Suyu},
  {Ding}, {Birrer}, {Chen}, {Shajib}, {Sluse}, {Wong}, {Agnello}, {Auger},
  {Buckley-Geer}, {Chan}, {Collett}, {Fassnacht}, {Hilbert}, {Koopmans},
  {Motta}, {Mukherjee}, {Rusu}, {Sonnenfeld}, {Spiniello}, \& {Van de
  Vyvere}}]{Millon2020a}
{Millon}, M., {Galan}, A., {Courbin}, F., {et~al.} 2020, \aap, 639, A101

\bibitem[{{M{\"o}ller} \& {Blain}(2001)}]{Moller2001}
{M{\"o}ller}, O. \& {Blain}, A.~W. 2001, \mnras, 327, 339

\bibitem[{{Oguri} {et~al.}(2012){Oguri}, {Inada}, {Strauss}, {Kochanek},
  {Kayo}, {Shin}, {Morokuma}, {Richards}, {Rusu}, {Frieman}, {Fukugita},
  {Schneider}, {York}, {Bahcall}, \& {White}}]{oguri2012}
{Oguri}, M., {Inada}, N., {Strauss}, M.~A., {et~al.} 2012, \aj, 143, 120

\bibitem[{{Paraficz} \& {Hjorth}(2009)}]{Paraficz2009}
{Paraficz}, D. \& {Hjorth}, J. 2009, \aap, 507, L49

\bibitem[{{Pascale} {et~al.}(2024){Pascale}, {Frye}, {Pierel}, {Chen}, {Kelly},
  {Cohen}, {Windhorst}, {Riess}, {Kamieneski}, {Diego}, {Meena}, {Cha},
  {Oguri}, {Zitrin}, {Jee}, {Foo}, {Leimbach}, {Koekemoer}, {Conselice}, {Dai},
  {Goobar}, {Siebert}, {Strolger}, \& {Willner}}]{Pascal2024}
{Pascale}, M., {Frye}, B.~L., {Pierel}, J. D.~R., {et~al.} 2024, arXiv
  e-prints, arXiv:2403.18902

\bibitem[{{Refsdal}(1964)}]{Refsdal1964}
{Refsdal}, S. 1964, \mnras, 128, 307

\bibitem[{{Rusu} {et~al.}(2017){Rusu}, {Fassnacht}, {Sluse}, {Hilbert}, {Wong},
  {Huang}, {Suyu}, {Collett}, {Marshall}, {Treu}, \& {Koopmans}}]{Rusu2017}
{Rusu}, C.~E., {Fassnacht}, C.~D., {Sluse}, D., {et~al.} 2017, \mnras, 467,
  4220

\bibitem[{{Schmidt} {et~al.}(2023){Schmidt}, {Treu}, {Birrer}, {Shajib},
  {Lemon}, {Millon}, {Sluse}, {Agnello}, {Anguita}, {Auger-Williams},
  {McMahon}, {Motta}, {Schechter}, {Spiniello}, {Kayo}, {Courbin}, {Ertl},
  {Fassnacht}, {Frieman}, {More}, {Schuldt}, {Suyu}, {Aguena},
  {Andrade-Oliveira}, {Annis}, {Bacon}, {Bertin}, {Brooks}, {Burke}, {Carnero
  Rosell}, {Carrasco Kind}, {Carretero}, {Conselice}, {Costanzi}, {da Costa},
  {Pereira}, {De Vicente}, {Desai}, {Doel}, {Everett}, {Ferrero}, {Friedel},
  {Garc{\'\i}a-Bellido}, {Gaztanaga}, {Gruen}, {Gruendl}, {Gschwend},
  {Gutierrez}, {Hinton}, {Hollowood}, {Honscheid}, {James}, {Kuehn}, {Lahav},
  {Menanteau}, {Miquel}, {Palmese}, {Paz-Chinch{\'o}n}, {Pieres}, {Plazas
  Malag{\'o}n}, {Prat}, {Rodriguez-Monroy}, {Romer}, {Sanchez}, {Scarpine},
  {Sevilla-Noarbe}, {Smith}, {Suchyta}, {Tarle}, {To}, {Varga}, \& {DES
  Collaboration}}]{Schmidt2023}
{Schmidt}, T., {Treu}, T., {Birrer}, S., {et~al.} 2023, \mnras, 518, 1260

\bibitem[{{Schneider}(2014{\natexlab{a}})}]{Schneider2014a}
{Schneider}, P. 2014{\natexlab{a}}, \aap, 568, L2

\bibitem[{{Schneider}(2014{\natexlab{b}})}]{Schneider2014b}
{Schneider}, P. 2014{\natexlab{b}}, arXiv e-prints, arXiv:1409.0015

\bibitem[{{Schneider} \& {Sluse}(2013)}]{Schneider2013}
{Schneider}, P. \& {Sluse}, D. 2013, \aap, 559, A37

\bibitem[{{Schuldt} {et~al.}(2019){Schuldt}, {Chiriv{\`\i}}, {Suyu},
  {Y{\i}ld{\i}r{\i}m}, {Sonnenfeld}, {Halkola}, \& {Lewis}}]{Schuldt2019}
{Schuldt}, S., {Chiriv{\`\i}}, G., {Suyu}, S.~H., {et~al.} 2019, \aap, 631, A40

\bibitem[{{Shajib} {et~al.}(2020){Shajib}, {Birrer}, {Treu}, {Agnello},
  {Buckley-Geer}, {Chan}, {Christensen}, {Lemon}, {Lin}, {Millon}, {Poh},
  {Rusu}, {Sluse}, {Spiniello}, {Chen}, {Collett}, {Courbin}, {Fassnacht},
  {Frieman}, {Galan}, {Gilman}, {More}, {Anguita}, {Auger}, {Bonvin},
  {McMahon}, {Meylan}, {Wong}, {Abbott}, {Annis}, {Avila}, {Bechtol}, {Brooks},
  {Brout}, {Burke}, {Carnero Rosell}, {Carrasco Kind}, {Carretero},
  {Castander}, {Costanzi}, {da Costa}, {De Vicente}, {Desai}, {Dietrich},
  {Doel}, {Drlica-Wagner}, {Evrard}, {Finley}, {Flaugher}, {Fosalba},
  {Garc{\'\i}a-Bellido}, {Gerdes}, {Gruen}, {Gruendl}, {Gschwend}, {Gutierrez},
  {Hollowood}, {Honscheid}, {Huterer}, {James}, {Jeltema}, {Krause},
  {Kuropatkin}, {Li}, {Lima}, {MacCrann}, {Maia}, {Marshall}, {Melchior},
  {Miquel}, {Ogando}, {Palmese}, {Paz-Chinch{\'o}n}, {Plazas}, {Romer},
  {Roodman}, {Sako}, {Sanchez}, {Santiago}, {Scarpine}, {Schubnell}, {Scolnic},
  {Serrano}, {Sevilla-Noarbe}, {Smith}, {Soares-Santos}, {Suchyta}, {Tarle},
  {Thomas}, {Walker}, \& {Zhang}}]{Shajib2020}
{Shajib}, A.~J., {Birrer}, S., {Treu}, T., {et~al.} 2020, \mnras, 494, 6072

\bibitem[{{Shajib} {et~al.}(2023){Shajib}, {Mozumdar}, {Chen}, {Treu},
  {Cappellari}, {Knabel}, {Suyu}, {Bennert}, {Frieman}, {Sluse}, {Birrer},
  {Courbin}, {Fassnacht}, {Villafa{\~n}a}, \& {Williams}}]{Shajib2023}
{Shajib}, A.~J., {Mozumdar}, P., {Chen}, G. C.~F., {et~al.} 2023, \aap, 673, A9

\bibitem[{{Shajib} {et~al.}(2021){Shajib}, {Treu}, {Birrer}, \&
  {Sonnenfeld}}]{Shajib2021}
{Shajib}, A.~J., {Treu}, T., {Birrer}, S., \& {Sonnenfeld}, A. 2021, \mnras,
  503, 2380

\bibitem[{{Sharma} {et~al.}(2023){Sharma}, {Collett}, \& {Linder}}]{Sharma2023}
{Sharma}, D., {Collett}, T.~E., \& {Linder}, E.~V. 2023, \jcap, 2023, 001

\bibitem[{{Sheu} {et~al.}(2024){Sheu}, {Shajib}, {Treu}, {Sonnenfeld},
  {Birrer}, {Cappellari}, {Oldham}, \& {Tan}}]{Sheu2024}
{Sheu}, W., {Shajib}, A.~J., {Treu}, T., {et~al.} 2024, arXiv e-prints,
  arXiv:2408.10316

\bibitem[{{Tanaka} {et~al.}(2016){Tanaka}, {Wong}, {More}, {Dezuka}, {Egami},
  {Oguri}, {Suyu}, {Sonnenfeld}, {Higuchi}, {Komiyama}, {Miyazaki}, {Onoue},
  {Oyamada}, \& {Utsumi}}]{Tanaka2016}
{Tanaka}, M., {Wong}, K.~C., {More}, A., {et~al.} 2016, \apjl, 826, L19

\bibitem[{{Tihhonova} {et~al.}(2020){Tihhonova}, {Courbin}, {Harvey},
  {Hilbert}, {Peel}, {Rusu}, {Fassnacht}, {Bonvin}, {Marshall}, {Meylan},
  {Sluse}, {Suyu}, {Treu}, \& {Wong}}]{Tihhonova2020}
{Tihhonova}, O., {Courbin}, F., {Harvey}, D., {et~al.} 2020, \mnras, 498, 1406

\bibitem[{{Tihhonova} {et~al.}(2018){Tihhonova}, {Courbin}, {Harvey},
  {Hilbert}, {Rusu}, {Fassnacht}, {Bonvin}, {Marshall}, {Meylan}, {Sluse},
  {Suyu}, {Treu}, \& {Wong}}]{Tihhonova2018}
{Tihhonova}, O., {Courbin}, F., {Harvey}, D., {et~al.} 2018, \mnras, 477, 5657

\bibitem[{{Tu} {et~al.}(2009){Tu}, {Gavazzi}, {Limousin}, {Cabanac},
  {Marshall}, {Fort}, {Treu}, {P{\'e}llo}, {Jullo}, {Kneib}, \&
  {Sygnet}}]{Tu2009}
{Tu}, H., {Gavazzi}, R., {Limousin}, M., {et~al.} 2009, \aap, 501, 475

\bibitem[{van Dokkum {et~al.}(2023)van Dokkum, Brammer, Wang, Leja, \&
  Conroy}]{vanDokkum2023}
van Dokkum, P., Brammer, G., Wang, B., Leja, J., \& Conroy, C. 2023, Nature
  Astronomy, 8, 119–125

\bibitem[{{Wells} {et~al.}(2023){Wells}, {Fassnacht}, \& {Rusu}}]{Wells2023}
{Wells}, P., {Fassnacht}, C.~D., \& {Rusu}, C.~E. 2023, \aap, 676, A95

\bibitem[{{Werner} {et~al.}(2008){Werner}, {An}, \& {Evans}}]{Werner2008}
{Werner}, M.~C., {An}, J., \& {Evans}, N.~W. 2008, \mnras, 391, 668

\bibitem[{{Wong} {et~al.}(2020){Wong}, {Suyu}, {Chen}, {Rusu}, {Millon},
  {Sluse}, {Bonvin}, {Fassnacht}, {Taubenberger}, {Auger}, {Birrer}, {Chan},
  {Courbin}, {Hilbert}, {Tihhonova}, {Treu}, {Agnello}, {Ding}, {Jee},
  {Komatsu}, {Shajib}, {Sonnenfeld}, {Blandford}, {Koopmans}, {Marshall}, \&
  {Meylan}}]{Wong2020}
{Wong}, K.~C., {Suyu}, S.~H., {Chen}, G. C.~F., {et~al.} 2020, \mnras, 498,
  1420

\bibitem[{{Wong} {et~al.}(2014){Wong}, {Tran}, {Suyu}, {Momcheva}, {Brammer},
  {Brodwin}, {Gonzalez}, {Halkola}, {Kacprzak}, {Koekemoer}, {Papovich}, \&
  {Rudnick}}]{Wong2014}
{Wong}, K.~C., {Tran}, K.-V.~H., {Suyu}, S.~H., {et~al.} 2014, \apjl, 789, L31

\end{thebibliography}

\end{document}